\documentclass[twocolumn,twocolappendix,trackchanges]{aastex7}
\usepackage{booktabs}
\usepackage{graphicx}
\usepackage{subcaption} 
\usepackage{adjustbox}  

\received{August, 2025}
\accepted{November, 2025}
\submitjournal{ApJ}

\begin{document}

\title{Constraining the Jet Energetics of the Transient X-ray Binaries MAXI J1348$-$630 and MAXI J1820$+$070 through Calorimetry}

\author[orcid=0000-0002-1514-5558,sname='Bosch-Cabot']{Pau Bosch-Cabot}
\affiliation{Department of Physics and Astronomy, University of Lethbridge, Lethbridge, Alberta, T1K 3M4, Canada}
\email[show]{pau.boschcabot@uleth.ca}  

\author[orcid=0000-0003-3906-4354,gname=Alexandra, sname='Tetarenko']{Alexandra J. Tetarenko} 
\affiliation{Department of Physics and Astronomy, University of Lethbridge, Lethbridge, Alberta, T1K 3M4, Canada}
\email{}

\author[orcid=0000-0002-5204-2259]{Erik Rosolowsky}
\affiliation{Department of Physics, University of Alberta, 4-183 CCIS, Edmonton, Alberta, T6G 2E1, Canada }
\email{rosolowsky@ualberta.ca}

\author[orcid=0000-0002-0426-3276]{Francesco Carotenuto}
\affiliation{INAF-Osservatorio Astronomico di Roma, Via Frascati 33, I-00078, Monte Porzio Catone (RM), Italy}
\email{}

\author[orcid=0000-0003-3124-2814]{James Miller-Jones} 
\affiliation{International Centre for Radio Astronomy Research, Curtin University, GPO Box U1987, Perth, WA 6845, Australia}
\email{}

\author[orcid=0000-0002-3500-631X ]{David M. Russell} 
\affiliation{Center for Astrophysics and Space Science, New York University Abu Dhabi, PO Box 129188, Abu Dhabi, UAE}
\email{}

\author[orcid=0000-0001-5538-5831 ]{Stéphane Corbel} 
\affiliation{Université Paris Cité and Université Paris Saclay,CEA, CNRS, AIM, F-91190 Gif-sur-Yvette, France}
\email{}

\author[orcid=0000-0002-7930-2276 ]{Thomas D. Russell} 
\affiliation{INAF, Istituto di Astrofisica Spaziale e Fisica Cosmica, Via U. La Malfa 153, I-90146 Palermo, Italy}
\email{}

\author[orcid=0000-0001-6682-916X]{Gregory R. Sivakoff}
\affiliation{Department of Physics, University of Alberta, 4-183 CCIS, Edmonton, Alberta, T6G 2E1, Canada }
\email{sivakoff@ualberta.ca}

\begin{abstract}

We present Atacama Large Millimeter/Submillimeter Array (ALMA) observations aimed at identifying potential jet-ISM interaction sites in the vicinity of the transient black hole X-ray binaries MAXI J1348$-$630 and MAXI J1820+070, both of which have recently undergone an outburst, and displayed powerful large scale jets. Using this dataset, we construct molecular line emission maps. By analyzing the morphological, spectral, and kinematic properties of the detected emission, we identify a molecular structure that provides compelling evidence for a jet-driven cavity in the local environment of MAXI J1348$-$630 but find no significant emission in the local environment of MAXI J1820$+$070. We use the properties of the detected molecular emission surrounding MAXI J1348$-$630 to constrain the jet power, finding our results to be consistent with other independent studies of this source, and further validating the utility of astrochemistry for constraining jet energetics. Additionally, our findings provide the first assessment on the formation timescales for jet-ISM interaction regions in the transient black hole X-ray binary population.


\end{abstract}



\section{Introduction}
\label{sec:Intr}
Black holes (BHs) are well-established as key regulators of their environments through the emission of powerful relativistic jets. These structures are related to the accretion dynamics of the source, which, in turn, means that they are also inherently bound to the nature of the BH. Jets, relativistic and collimated in nature, carry substantial kinetic energy away from the accretion flow and into the environment. However, the characterization of the jet's energetics and matter content remains a fundamental challenge in the field \citep{fender2015overview}.

When it comes to active galactic nuclei (AGN), it has been well established that the radiative and mechanical feedback provided by jets plays a crucial role in the co-evolution of super-massive BHs (SMBHs) and their environments, injecting energy on scales sufficient to influence the star formation dynamics and galaxy evolution in significant ways \citep[e.g.,][]{magorrian1998demography, mcnamara2005heating, mcnamara2007heating, mirabel2011stellar}. This co-evolution allows to establish a link between the jet properties observed and the physical properties of the underlying BH. Particularly, there has been indication that the jet power is related to the BH spin \citep{blandford1977electromagnetic} even if more complicated processes also play a significant role into jet properties \citep{blandford1982hydromagnetic}. Therefore, finding an efficient way to probe jet power is of crucial importance in the field.

Stellar-mass black holes also exhibit powerful interactions with their surrounding interstellar medium (ISM), establishing a crucial link between the physics of accretion and that of the ISM. Galactic black hole X-ray binaries (BHXBs) and ultra-luminous x-ray sources (ULXs)—highly accreting systems in nearby galaxies with X-ray luminosities near the Eddington limit ( $L_X <10^{39}$erg s$^{-1}$, \citealt{kaaret2017ultraluminous})—are known to deposit considerable energy into their surroundings, frequently displaying jet-blown cavities or ionized nebular structures surrounding the compact object \citep[e.g.,][]{corbel2002large, heinz2005estimating, gallo2005dark, fender2005energization, russell2010powerful, soria2010radio, pakull2010300, cseh2014unveiling, tetarenko2018mapping, tetarenko2020jet, carotenuto2022modelling}.

Galactic BHXBs are prime laboratories for studying jet energetics. This is due to the fact that they display short evolutionary timescales in their transient cycles (ranging from hours to months), proximity, and scale-invariant similarities to AGN jets. Over the past two decades, numerous observational surveys—enabled by increasing instrumental sensitivity—have identified potential jet-ISM interaction sites within the Milky Way and the Magellanic Clouds, including SS 433 \citep{dubner1998high}, Cygnus X--1 \citep{gallo2005dark, russell2007jet, sell2015shell}, GRS 1758--258 \citep{mirabel1992double}, GRS 1915$+$105 \citep{kaiser2004revision, rodriguez1998surroundings, chaty2001search, motta2025meerkat, tetarenko2018mapping}, H1743--322 \citep{corbel2005discovery}, XTE J1550--564 \citep{corbel2002large, kaaret2003x, migliori2017evolving}, XTE J1748--288 \citep{brocksopp2007highly}, GRO J1655--40 \citep{hjellming1995episodic, hannikainen2000radio}, GX 339--4 \citep{gallo2004transient}, 4U 1755--33 \citep{kaaret2006evolution}, XTE J1752--223 \citep{yang2010decelerating, yang2011transient, miller2011accurate, ratti2012black}, XTE J1650--500 \citep{corbel2004origin}, XTE J1908$+$094 \citep{rushton2017resolved}, 4U 1630--47 \citep{neilsen2014link, kalemci2018dust}, LMC X--1 \citep{russell2006jet, cooke2007spectacular, hyde2017lmc}, and GRS 1009--45 \citep{russell2006jet}.

A comprehensive understanding of the morphology and composition of these jet-ISM interaction sites offers a unique opportunity to constrain jet properties that are otherwise notoriously challenging to measure, such as jet power, velocity, and composition \citep[e.g.,][]{mcnamara2007heating, burbidge1959estimates, castor1975interstellar, heinz2006composition}. These properties are fundamental to understanding the mechanisms governing jet formation and acceleration. However, direct measurement has historically been difficult, often leading to reliance on broad theoretical assumptions that impact the interpretation of observational data. Notably, identifying jet-ISM interaction structures can offer the possibility of using them as calorimeters to infer the minimum jet power required to shape and sustain them, thus reducing dependence on the aforementioned theoretical assumptions \citep[e.g.,][]{gallo2005dark, russell2007jet, sell2015shell, tetarenko2018mapping, tetarenko2020jet}.

However, characterizing jet-ISM interaction sites presents even more challenges due to multiple factors. In particular, environmental parameters such as gas density, kinetic temperature, and shock velocity --all of which play crucial roles in the calorimetric approach-- are impossible to constrain via continuum imaging alone \citep[e.g.,][]{phillips1994images,russell2007jet, sell2015shell}. These limitations underscore the need for alternative observational strategies capable of providing a more comprehensive characterization of the jet-ISM interaction. 
Given the substantial energy injection by jets into the ISM, significant alterations in the chemical and excitation conditions of the affected gas are expected at interaction sites. Taking advantage of these conditions, an opportunity arises: characterizing the properties of the ISM near the source via several molecular tracers to establish a direct connection between the measurable ISM properties and those of the jet \citep{tetarenko2018mapping, tetarenko2020jet}. Molecular line imaging enables simultaneous investigation of the physical and chemical conditions of the ISM while preserving structural information about any potential jet-formed cavities.
The Atacama Large Millimeter/submillimeter Array (ALMA) offers the bandwidth and sensitivity required to simultaneously track these molecular line tracers in a single observation, significantly improving the efficiency and scalability of this method compared to previous studies.

Following two successful applications of this astrochemistry technique to BHXBs that persistently accrete and launch jets continuously over decades-long timescales \citep{tetarenko2018mapping, tetarenko2020jet}, we extend this methodology to transient BHXBs, which enter short-lived (months--long) bright outbursts. In particular, we target the black hole X-ray binaries MAXI J1820$+$070 and MAXI J1348--630, both of which recently underwent bright outburst periods and show tantalizing evidence for jet-ISM interactions in the form of decelerating jet motion \citep{carotenuto2022modelling,espinasse2020relativistic, bright2020extremely}. 
This is the first time such an approach has been carried out for transient sources, further providing a first glimpse at the timescales involved in the formation of such jet-driven structures in the ISM. By employing molecular line imaging, we aim to identify jet-ISM interaction sites with a high degree of confidence and place robust constraints on key jet properties, thereby advancing our understanding of BH energetics and feedback processes.

In \S\ref{sec:Obs}, we describe the details of the observations and data reduction procedures. In \S\ref{sec:Res}, we present the results derived from our analysis, while in \S\ref{sec:Disc}, we discuss the implications of our findings and highlight areas of future study.

\begin{table*}[t!]
    \caption{ALMA spectral setup. Frequencies are in reference to the Kinematic Local Standard of Rest (LSRK).}
    \centering
    \renewcommand{\arraystretch}{1.2}
    \begin{tabular}{c c c c c l}
        \toprule
        \textbf{Array} & \textbf{Target Line} & \textbf{Rest LSRK} & \textbf{ACA+12m} & \textbf{ACA+12m} & \textbf{What does it trace?}  \\ \textbf{Config.}
        & & \textbf{Freq. (GHz)} & \textbf{Bandwidth} &\textbf{Resolution} & \\
        & & & (km s$^{-1}$) & (km s$^{-1}$) & \\
        \midrule
        B & HCN ($\nu = 0, J = 1 - 0$) & 88.631601 & 792.7 & 1.321 &Density ($n\sim10^4-10^5$cm$^{-3}$) \\
        B & HCO$^+$ ($\nu = 0, J = 1 - 0$) & 89.18853 & 787.8 & 1.313 &Density ($n\sim10^4-10^5$cm$^{-3}$),\\
        &&&&&ionization \\
        B & SiO ($\nu = 0, J = 2 - 1$) & 86.84696 & 809 & 1.348 &Shocks\\
        B & HNCO ($\nu = 0, J = 4_{0,4} - 3_{0,3}$) & 87.92524 & 799.1 & 1.332 &Density ($n\sim10^5-10^7$cm$^{-3}$),\\
        &&&&&shocks, temperature\\
        B & CS ($\nu = 0, J = 2 - 1$) & 97.98095 & 358.5 & 1.195 &Density($n\sim10^4-10^5$cm$^{-3}$),\\
        &&&&&shocks.\\
        A & HNCO ($\nu = 0, J = 5_{0,5} - 4_{0,4}$) & 109.90575 & 319.7 & 1.066 &Density ($n\sim10^5-10^7$cm$^{-3}$),\\
        &&&&&shocks, temperature\\
        A & C$^{18}$O ($\nu = 0, J = 1 - 0$) & 109.78218 & 320 & 1.067 &Density ($n\sim10^2-10^3$cm$^{-3}$),\\
        &&&&&opacity\\
        A & $^{13}$CO ($\nu = 0, J = 1 - 0$) & 110.20135 & 318.8 & 1.063 &Density ($n\sim10^2-10^3$cm$^{-3}$),\\
        &&&&&opacity \\
        A,B & Continuum$^\dagger$ & 96.0 & 5885.4 & 130.121 & Synchrotron jet emission \\
         &  & 97.5 & 5763.3 & 128.119 & \\
         &  & 100.0 & 5620.8 & 124.906 & \\
        & & 108.0 & 5204.8 & 115.663 &  \\
        \bottomrule
    \end{tabular}
    \\
    \textsuperscript{$\dagger$}We refer to the central frequency in the spectral window for each continuum observation.
    \label{tab:config}
\end{table*}

\subsection{MAXI J1348$-$630}
MAXI J1348$-$630 is a BHXB that was first detected in outburst by the Monitor of All-sky X-ray Image (MAXI) instrument aboard the International Space Station (ISS; \citealt{matsuoka2009maxi}) in 2019 \citep{yatabe2019maxi}, and it was subsequently identified as BH candidate \citep{kennea2019maxi,zhang2020nicer}. The system hosts a BH whose mass has not been inferred through dynamical mass measurements yet. There have been several attempts to measure the BH spin, defined by the dimensionless spin parameter $a_*$, constraining the value to $a_*\sim 0.79-0.97$ \citep{2022MNRAS.513.4869K, mall2022broadband, jia2022detailed, guan2021physical}. The jet inclination angle with respect to the line of sight remains unconstrained, discarding only very large angles \citep[$i\lesssim80^\circ$]{cooper2025joint}.
The distance to MAXI J1348--630 has been estimated through two independent methods: \citet{chauhan2021measuring} derived a kinematic distance of $2.2^{+0.5}_{-0.6}$ kpc based on HI absorption features, whereas \citet{lamer2021giant} used XMM-Newton observations of a giant scattering dust ring to obtain a distance of $3.39 \pm 0.34$ kpc. Radio wavelength studies tracking jet motion through direct imaging during this outburst were suggestive of a surrounding environment that significantly decelerates the jets over time \citep{carotenuto2021black}. Modeling by \cite{carotenuto2022modelling} inferred the presence of a potential jet-driven cavity in the local environment with a radius after observing the ejecta decelerate quickly at an angular distance from the source around $\sim25''$. 
The article also set constrains for the kinetic energy of the ejecta that were later revised by \citealt{zdziarski2023no}. This later work suggests the kinetic energy to be $\log_{10}(E_0/\text{erg})\sim 43.7-44.5$.

\subsection{MAXI J1820$+$070}
MAXI J1820+070 is a BHXB that was first detected in the optical band as a transient by the ASAS-SN survey and later detected in the X-ray band by MAXI during an outburst in 2018 \citep{tucker2018asassn, shidatsu2018x}. The system hosts a BH with a mass of $M \simeq (5.95 \pm 0.22) M_\odot / \sin^3 i_b$, where $i_b$ is the inclination of the binary orbital plane with respect to the line of sight \citep{torres2020binary}. Several attempts have been made to determine the BH spin, constraining the value to $a_* \sim 0.2 - 0.8$ \citep{guan2021physical, zhao2021estimating, bhargava2021timing}. Radio parallax methods establish the distance to the source, $D \simeq 2.96 \pm 0.33$ kpc \citep{atri2020radio}. The inclination angle of the orbital plane relative to the line of sight was initially constrained to be $i_b \sim 66^\circ - 88^\circ$ \citep{torres2020binary}, and later refined through radio imaging of the jet to be $i_b \simeq 64^\circ \pm 5^\circ$ \citep{wood2021varying}. Similar to MAXI J1348--630, radio and X-ray studies tracking jet motion through direct imaging showed decelerating jets over time, suggesting a possible interaction with the surrounding ISM \citep{espinasse2020relativistic, bright2020extremely}.

\begin{figure*}[t!]
    \centering
    \begin{subfigure}[b]{0.52\textwidth} 
        \centering
        \includegraphics[width=\textwidth]{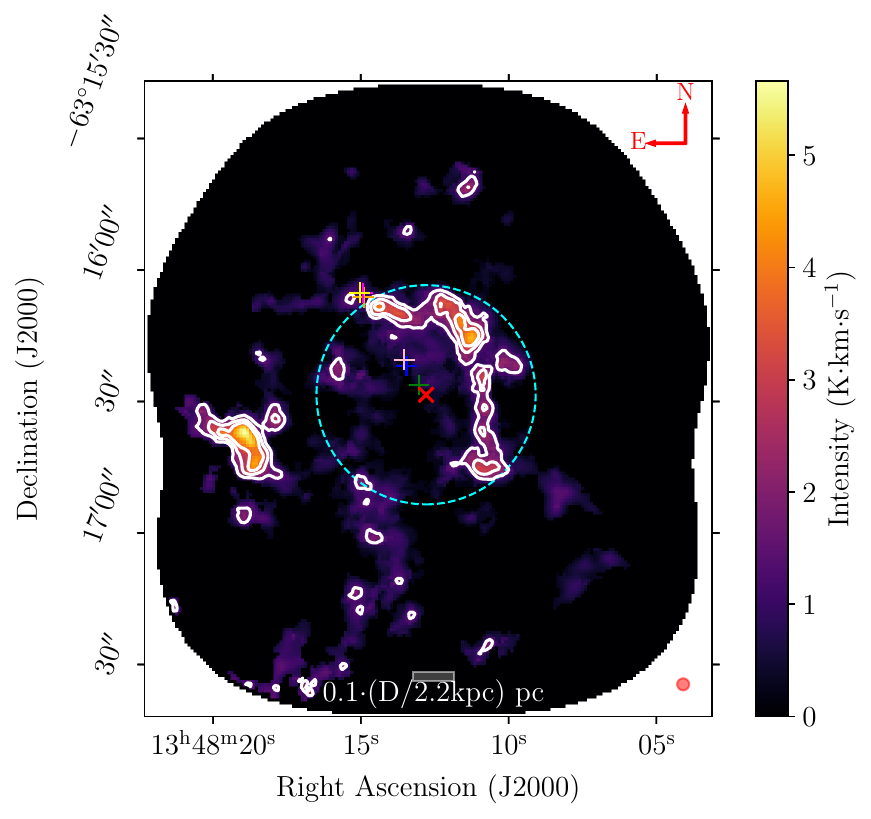}

    \end{subfigure}
    \hfill
    \begin{subfigure}[b]{0.45\textwidth} 
        \centering
        \includegraphics[width=\textwidth]{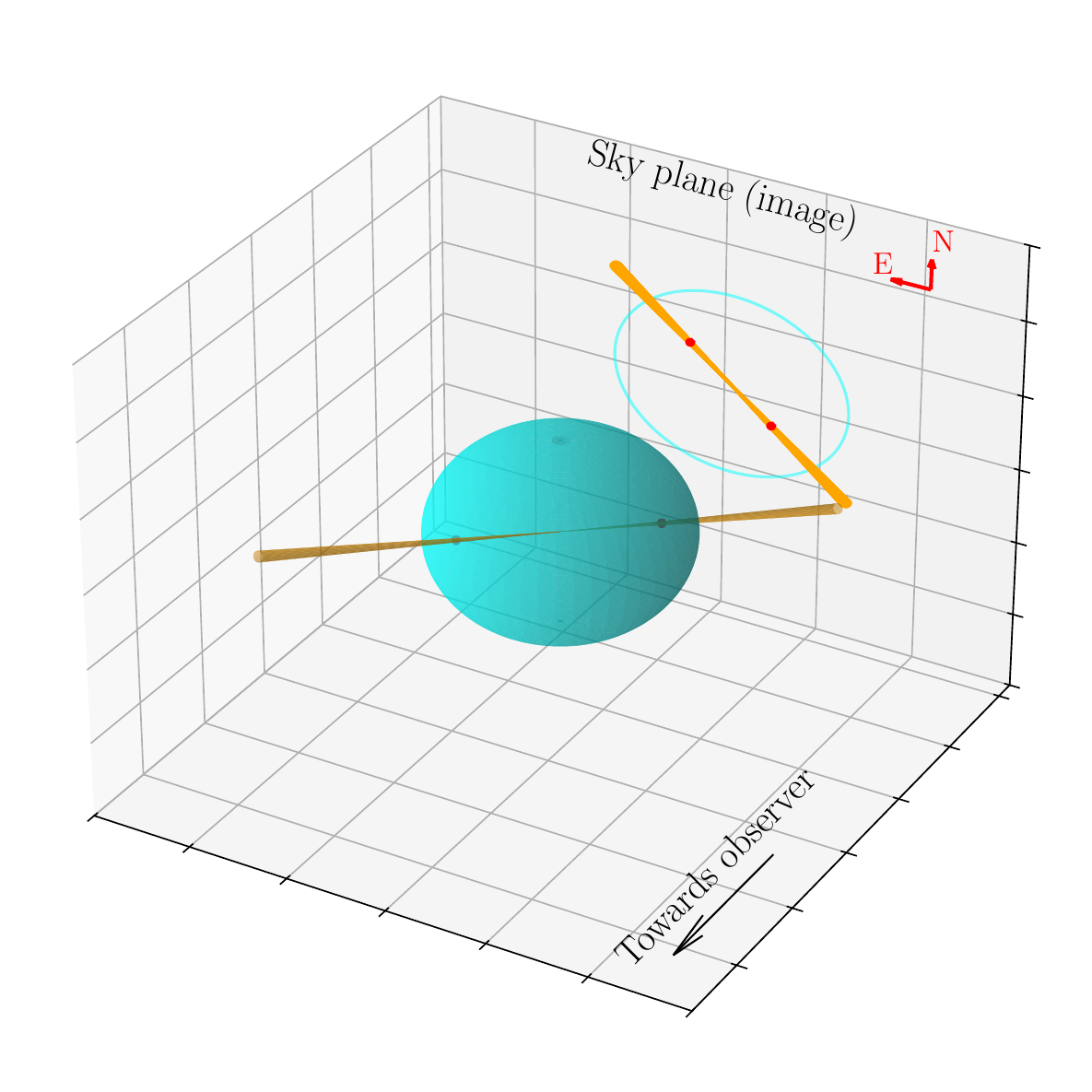}

    \end{subfigure}
    \caption{\textit{Left:} Integrated and $4\sigma$-filtered $^{13}$CO($J=1-0$) intensity map of the MAXI J1348$-$630 field, combining ALMA 12m and ACA data. Contours are at [1.5, 2.5, 3.5] K km s$^{-1}$. Dashed cyan lines mark the radius where the ejecta decelerate; the BHXB position is shown by a red cross, and ejecta locations observed with MeerKAT and ATCA by colored crosses. The red circle indicates the synthesized beam. \textit{Right:} Schematic linking the 2D emission to the 3D jet–ISM geometry: jets (orange cones), impact sites (red dots), and a possible cavity (cyan ellipsoid). The sketch is illustrative only—angles are not to scale. The ejecta slow near bright molecular emission, and the inferred cavity matches a ring-like structure.}
    \label{fig:main}
\end{figure*}

\section{Observations and data analysis} \label{sec:Obs}
\subsection{ALMA sub-mm observations}
We observed MAXI J1348$-$630 and MAXI J1820$+$070 with ALMA under project code 2023.1.00983.S. These observations were conducted using the 12m array on January 18 and April 21, 2024 (MJD 60303, MJD 60497) for MAXI J1348$-$630, and on January 13 and January 26, 2024 (MJD 60298, MJD 60311) for MAXI J1820+070. Additionally, we obtained observations with the Atacama Compact Array (ACA) on multiple dates: October 4, 2023, January 22, 26, and 27, 2024 (MJD 60232, MJD 60307, MJD 60311, MJD 60312) for MAXI J1348$-$630, and October 16, December 12, 14, 16, and 18, 2023 (MJD 60244, MJD 60301, MJD 60303, MJD 60305, MJD 60307) for MAXI J1820+070. All observations utilized the Band 3 receiver (84–-116 GHz).

For each source, we employed two spectral setups (denoted as A and B in Table~\ref{tab:config}), utilizing 43 antennas for the 12m array and 10 antennas for the ACA array. In the case of MAXI J1820+070, the A/B configurations in the 12m+ACA arrays included 4+3/7+3 pointings, respectively, centered on the source coordinates RA$=18^{\mathrm{h}}20^{\mathrm{m}}21.94^{\mathrm{s}}$, DEC$=+07^{\circ}11'07.28''$, covering fields of $60'' \times 60''$ in all cases. Similarly, for MAXI J1348$-$630, the A/B configurations in the 12+7m arrays consisted of 10+4/7+3 pointings centered on RA$=13^{\mathrm{h}}48^{\mathrm{m}}12.74^{\mathrm{s}}$, DEC$=-63^{\circ}16'29.71''$, with fields of $72'' \times 72''$. The field of view and angular resolution achieved with all the ALMA observations were designed to match the scales of previously observed radio/X-ray jet imaging \citep{carotenuto2022modelling,espinasse2020relativistic, bright2020extremely}.

The ALMA correlator was configured to produce $4\times 2$ GHz wide base bands, within which we defined 12 spectral windows centered on targeted molecular lines, alongside several continuum windows (see Table~\ref{tab:config}). All data reduction and imaging were carried out using the Common Astronomy Software Applications package (\textsc{casa}, version 6.6.1.17; \citealt{mcmullin2007casa}). The data was calibrated using the ALMA pipeline \citep{hunter2023alma} and imaged with the PHANGS--ALMA pipeline with a strict $4\sigma$ filter that suppresses the contribution of the noise\footnote{We use the standard PHANGS imaging strategy described in \citealt{leroy2021phangs}, which creates a mask requiring the detection of signal in two adjacent channels at $\geq4 \sigma$.} (for molecular line emission; \citealt{leroy2021phangs}) and the \texttt{tclean} task in multi-frequency synthesis mode for continuum imaging. We utilized a multiscale deconvolver (scales of \texttt{[0,1,2,5,10,20]} pixels) and natural weighting to maximize sensitivity in the imaging process. All images produced combined the 12m and ACA observations, to capitalize on both angular resolution and sensitivity, after having assessed imaging for the 12m and ACA data individually.

\begin{figure*}[t!]
    \centering
    \begin{subfigure}[b]{0.48\textwidth} 
        \centering
        \includegraphics[width=\textwidth]{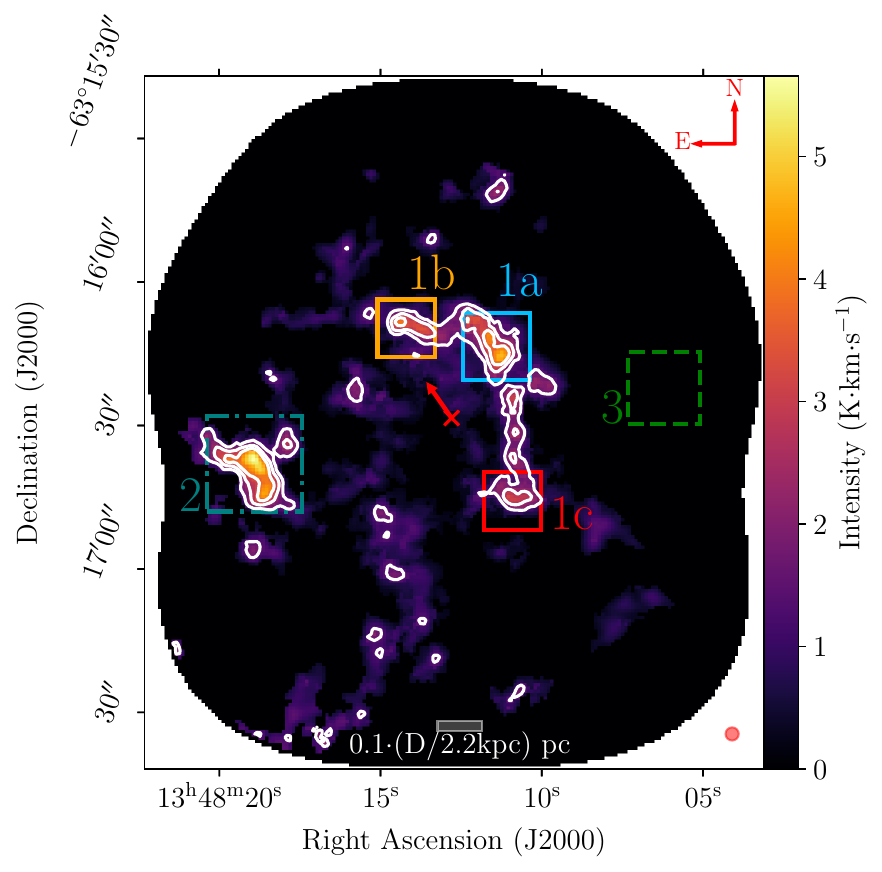}

    \end{subfigure}
    \hfill
    \begin{subfigure}[b]{0.50\textwidth} 
        \centering
        \includegraphics[width=\textwidth]{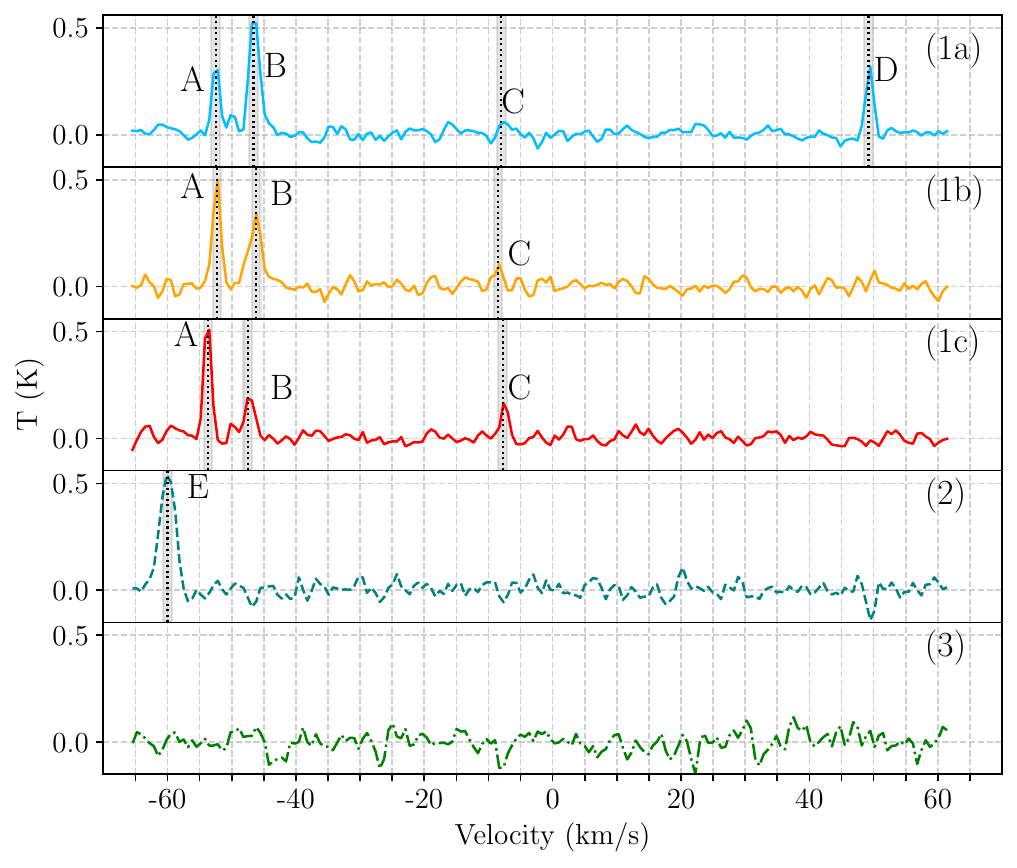}

    \end{subfigure}
    \caption{Spectral analysis of $^{13}$CO($J=1-0$) emission in the MAXI J1348$-$630 field. \textit{Left:} Integrated and $4\sigma$-filtered intensity map with contours at [1.5, 2.5, 3.5] K km s$^{-1}$. The BHXB position (red cross) and approaching jet direction (red arrow) are marked. Spectral extraction regions are shown: Regions 1a–c (blue, orange, red) trace the ring feature; Region 2 (teal) samples an isolated cloud; Region 3 (green) an off-emission area. The red circle indicates the synthesized beam. \textit{Right:} Corresponding spectra from the regions in the \textit{left} panel. Vertical black lines mark feature centroids with uncertainties (gray bands). Regions 1a–c show a double-peaked, asymmetric profile (features A and B), consistent with a jet–ISM interaction, while additional components (C–E) likely trace unrelated molecular gas along the line of sight.}
    \label{fig:spectra}
\end{figure*}

\section{Results} \label{sec:Res}
\subsection{Properties of the continuum emission}
No significant continuum emission was detected for either of the target sources analyzed in this study, regardless of the dataset utilized. This non-detection persisted across observations conducted with the 12 m array, the ACA array, and the combined 12 m + ACA dataset.
To maximize the signal recovered, we combined all of the continuum windows from both the 12m and ACA data, yielding a $3\sigma$ upper limit of continuum emission to be $F<60 \mu$Jy beam$^{-1}$ for both MAXI J1348$-$630 and MAXI J1820$+$070. 

In the case of MAXI J1348--630, previous mm-wavelength observations during outburst detected continuum emission between 12--18 mJy \citep{mandal2024multi}. During many phases of the outburst there were instances of non-detections in the radio monitoring \citep{carotenuto2021black}, with continuum emission levels found as low as 15$\mu$Jy (at 5.5GHz; \citealt{carotenuto2022black}).
In the case of MAXI J1820$+$070, previous mm-wavelength observations during outburst detected continuum emission between 2--120 mJy (at 230--343GHz; \citealt{tetarenko2018sub}). Therefore, in both cases, it is not expected to find mm-wavelength continuum emission around ALMA sensitivity levels at the time of our observations, which were taken a few years post outburst.

\begin{figure*}[t!]
    \centering
    \begin{minipage}{0.58\textwidth}
        \centering
        \includegraphics[width=\textwidth]{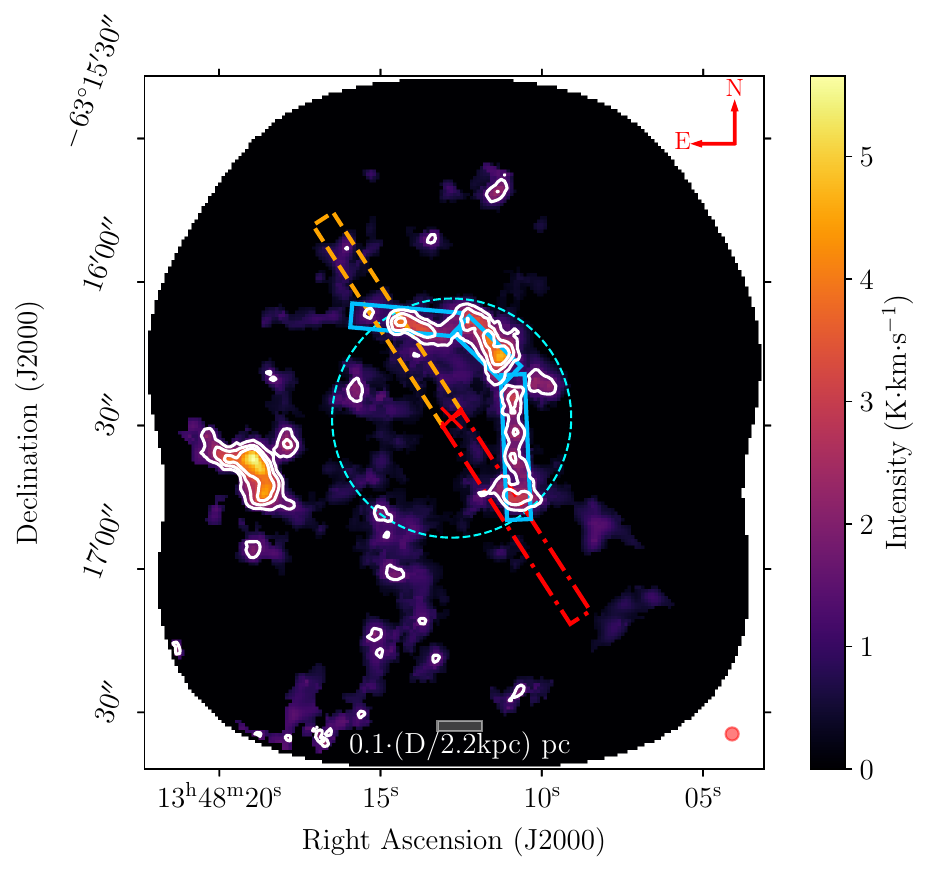}
    \end{minipage}
    \begin{minipage}{0.4\textwidth}
        \centering
        \begin{minipage}{\textwidth}
            \centering
            \includegraphics[width=\textwidth]{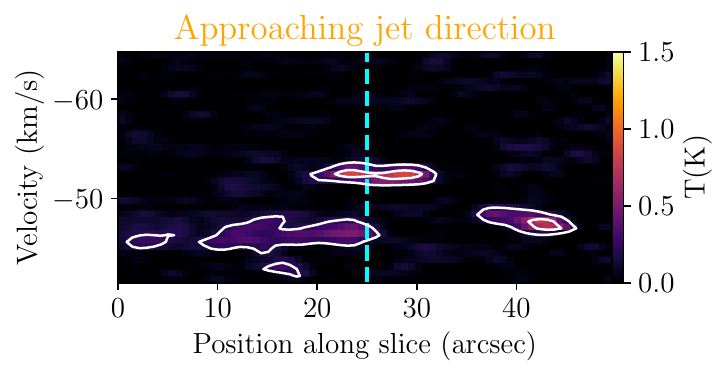}
        \end{minipage}
        
        \vspace{3pt} 
        
        \begin{minipage}{\textwidth}
            \centering
            \includegraphics[width=\textwidth]{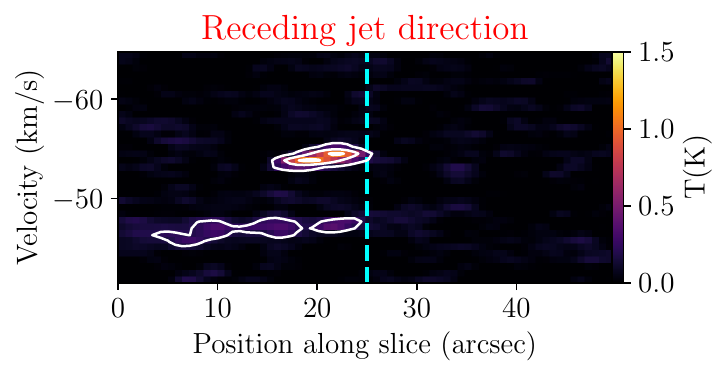}
        \end{minipage}
        
        \vspace{3pt} 
        
        \begin{minipage}{\textwidth}
            \centering
            \includegraphics[width=\textwidth]{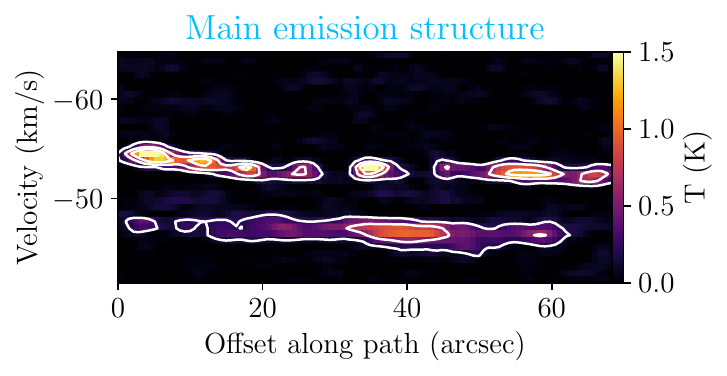}
        \end{minipage}
    \end{minipage}
    \caption{Kinematic analysis of $^{13}$CO($J=1-0$) emission in the MAXI J1348$-$630 field. \textit{Left:} Integrated and $4\sigma$-filtered intensity map with contours at [1.5, 2.5, 3.5] K km s$^{-1}$. The BHXB position is marked with a red cross, and the ejecta deceleration radius with a dashed cyan line. PV extraction paths are indicated: along the approaching jet (orange) and receding jet (red), both from the BHXB, and across the main ring emission (blue) from its southern end. The red circle shows the synthesized beam. \textit{Right:} Position–velocity diagrams along the slices in the \textit{left} panel, with contours at [0.2, 0.6, 1.0] K and the deceleration region marked by dashed cyan lines. Displaced gas is evident at the approaching/receding jet impact sites.}
    \label{fig:PV}
\end{figure*} 
\subsection{Properties of the molecular emission} \label{sec:molec}

Significant emission from 13 CO (J = 1$-$0) in the MAXI J1348$-$630 $72^{''} \times 72^{''}$ field was detected. However, MAXI J1820$+$070 did not show any significant molecular emission in the $60^{''} \times 60^{''}$ imaged field. We therefore focus the remainder of our analysis on the MAXI J1348$-$630 field in the $^{13}$CO spectral line.

In the MAXI J1348$-$630 field, we detect $^{13}$CO emission exhibiting a ring-like morphology surrounding the position of the target BHXB. This structure is consistent with the suspected jet blown cavity (see Fig.~\ref{fig:main}, regions 1a, 1b, 1c Fig.~\ref{fig:spectra}). The brightest molecular emission in this region appears to be aligned with the jet axis. In particular, the approaching jet is directed toward the NE, where the intersection of the jet axis and the observed molecular ring structure spatially coincides with the region where past jet motion observations show evidence of deceleration (i.e., the approaching jet impact site). Following the jet axis in the opposite direction, we find a symmetrical counterpart in molecular emission which could be attributed to the impact site of the unobserved receding jet (SW of the central BHXB). Both presumed impact sites lie along the cyan dashed line, giving indication that the cavity could have some degree of symmetry. Beyond the ring structure, we also find another bright, extended molecular cloud feature located much further from the BHXB (SE of the central BHXB, region 2 Fig.~\ref{fig:spectra}).

\begin{figure*}[t!]
    \centering
    \begin{subfigure}[b]{0.49\textwidth} 
        \centering
        \includegraphics[width=\textwidth]{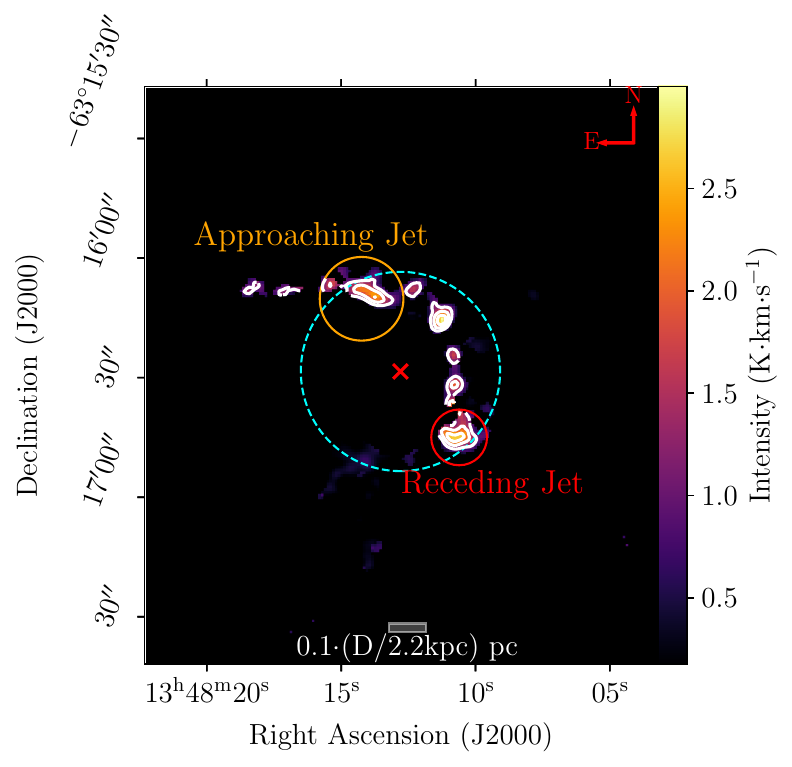}

    \end{subfigure}
    \hfill
    \begin{subfigure}[b]{0.49\textwidth} 
        \centering
        \includegraphics[width=\textwidth]{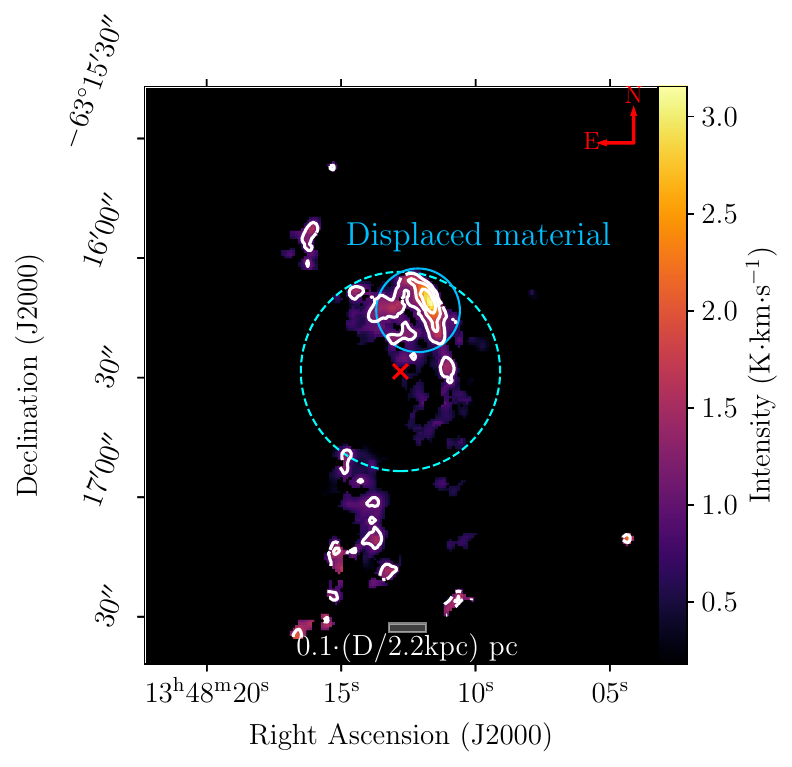}

    \end{subfigure}
    \caption{Integrated and $4\sigma$ filtered intensity maps of $^{13}$CO($J=1-0$) molecular emission in the MAXI J1348$-$630 field over specific velocity ranges. Contours correspond to [1.05, 1.75, 2.45]K km s$^{-1}$. The cyan dashed circles represent the radius at which the ejecta decelerate, while the red cross is the position of the target BHXB. The {\it left} panel corresponds to the velocity range $v=(-54,-50)$ km s$^{-1}$ (spectral feature A, Fig.~\ref{fig:spectra}, \textit{right}) and the {\it right} panel corresponds to the velocity range $v=(-49,-45)$ km s$^{-1}$ (spectral feature B, Fig.~\ref{fig:spectra}, \textit{right}). } 
    \label{fig:features}
\end{figure*}

To examine the properties of the identified bright molecular emission features, we first extracted the spectra in several of these regions (see Fig.~\ref{fig:spectra}); Region 1a, 1b, and 1c are located within the boundaries of the suspected jet-blown cavity along the ring structure, Region 2 covers the eastward bright extended cloud emission, and Region 3 represents an emission-free area for comparison. 
The spectra extracted from Regions 1a, 1b, and 1c exhibit double-peaked profiles (features A and B in Fig.~\ref{fig:spectra}, {\it right}), with varying intensities centered around $v_A = -52 \pm 2$ km s$^{-1}$ (feature A) and $v_B = -47 \pm 3$ km s$^{-1}$ (feature B). Such multi-peaked and asymmetric spectral features can arise from collisions between gas clouds moving at different velocities, and are a known indicator of jet-induced interactions \citep{tetarenko2018mapping, tetarenko2020jet}. The observed double-peaked structures likely reflect complex kinematic processes associated with this jet–ISM interaction. The peak intensity of these two spectral features varies between regions (feature A is stronger than feature B at the presumed jet impact sites of regions 1b and 1c). However, we find spectral feature B to be consistently wider spread than feature A. The central velocities show small shifts from region to region. The jet impact sites (1b,1c) display the most positive/negative velocities, with region 1a along the cavity structure lying in between the two. As we expect more negative velocities to indicate material flowing more towards the observer and more positive velocities to indicate material flowing more away from the observer, these small velocity shifts between regions 1b and 1c could be attributed to the motion of material rebounding back towards the central BHXB and along the cavity wall from the jet impact sites. Since the velocity shift is small ($\sim1$ km s$^{-1}$), the impact site molecular clouds (1b, 1c) might have very small peculiar velocities with respect to galactic rotation, and are certainly not displaying relativistic motion. We acknowledge that this is not the expected behavior for jet-boosted structures \citep{tetarenko2020jet}, where we would expect to see clear blue-shifting (more negative velocities) on the approaching side and clear red-shifting (more positive velocities) in the receding side. However, given the consistent size and alignment of these features, we attribute this to a very dense shocked medium that moved very little after the jet impact.


Additionally, we find other spectral features around $v_C=-8$ km s$^{-1}$ (feature C), $v_D=48$ km s$^{-1}$ (feature D, only in Region 1c), and $v_E=-61$ km s$^{-1}$ (feature E, only in Region 2).  Since they display very different central velocities compared to features A and B, we suspect these features likely track emission from unrelated molecular clouds found along the line of sight (see Appendix \ref{sec:otheremiss}). While these features could correspond to physically close locations to the source BHXB, the fact that these additional lines do not show double-peaked structure or asymmetric profiles makes it unlikely that they would be associated with the jet-ISM interaction.  
Lastly, the spectra sampled from Region 3 shows no discernible spectral line features, as expected from an off-emission region.
In \S\ref{sec:Disc} we further discuss the origin and relationship between these spectral features.

To better understand the kinematics of the identified bright molecular emission features, we also constructed position–velocity (PV) diagrams (see Fig.~\ref{fig:PV}), tracing the spectral profiles along specific paths across the emission regions, allowing us to better correlate the spectral and morphological properties of the emission. 
Along both the approaching and receding jet directions (orange and red line regions in Fig.~\ref{fig:PV}), we observe similar PV structures consistent with displaced gas (traced by a dominant spectral feature A; see also Fig.~\ref{fig:spectra}) located at the suspected cavity wall (cyan line in Fig.~\ref{fig:PV}). The fact that the structures are similar towards both directions points towards a likely first detection of the receding jet for MAXI J1348$-$630. Along the main ring structure, we observe a slight velocity gradient (more negative velocities on the receding jet side compared to more positive velocities on approaching jet side), and emission spanning a broader velocity range compared to the other two PV slices. The broader velocity range in this PV slice compared to the jet impact site PV slices could also indicate a more complex kinematic distribution of the gas in this region. Since the main emission around spectral feature A in this PV slice shows a smooth gradient, it suggests that the whole structure is related, with a complex kinematic structure.

Finally, after analyzing the spectra and the PV diagrams we imaged the molecular emission around two critical velocity ranges (see Fig.~\ref{fig:features}): 
\begin{itemize}
    \item In the range $v=(-54,-50)$ km s$^{-1}$ (around spectral feature A, Fig.~\ref{fig:spectra}, \textit{right}) we find a series of clumps among which we can highlight two larger regions. Due to their alignment with the jet axis and location coincident with the deceleration region, we believe they could be associated with the jet impact sites.
    \item In the range $v=(-49,-45)$ km s$^{-1}$ (around spectral feature B, Fig.~\ref{fig:spectra}, \textit{right}) we find a much more sparse distribution with a major region NW from the target. We interpret this region to be related to material displaced by the jet.
\end{itemize}


\section{Discussion} \label{sec:Disc}
\subsection{Distance constraints}
The validity of any proposed jet-ISM interaction relies on the consistency of the inferred distances to the observed molecular features with the known distance to the BHXB. To assess this, we applied the kinematic distance estimation methods outlined in \cite{wenger2018kinematic}, utilizing the Galactic rotation parameters from \cite{reid2019trigonometric}.
Constraints from independent distance measurements \citep{lamer2021giant,chauhan2021measuring} locate the source between 2.2 and 3.4 kpc. In Galactic rotation this corresponds to a velocity range between -20 km s$^{-1}$ and -55 km s$^{-1}$.
Spectral features A and B were found at velocities of $v_A = -52 \pm 2$ km s$^{-1}$ and $v_B = -47 \pm 3$ km s$^{-1}$, both of which lie within this kinematic distance range, and therefore are likely to be directly associated with the BHXB source.
Further, this BHXB lies within $2^\circ$ of the Galactic plane (at $l \sim 309^\circ$), where known molecular cloud velocities range from $-60$ km s$^{-1}$ to $+50$ km s$^{-1}$ \citep{dame2001milky}. This is consistent with our theory that the spectral features C, D and E are consistent with line-of-sight background or foreground molecular clouds unrelated to the BHXB.

\subsection{A Supernova Remnant?}
Given the fact that supernova remnants (SNR) are normally associated with the presence of compact objects, we need to assess whether the observed structure could be consistent with a supernova explosion. 

To do so, we take an approach typically used in this kind of structures (e.g., \citealt{sell2015shell}): the Sedov-Taylor self similar solution  for expanding blast waves \citep{taylor1950formation,sedov2018similarity}. We employ the constraints on density ($n\sim 100-5000$ cm$^{-3}$)
and size ($L_j \sim 0.3-1$ pc) based on our observations (introduced in \S \ref{sec:jetprop}) and assume a lifetime of 1 Gyr to estimate the energy of the explosion needed to produce such a structure. We find that the estimated energy has an upper limit of $\lesssim10^{39}$ erg, well under the $10^{51}$ erg energies typically associated to supernovae explosions (e.g., \citealt{chevalier1977interaction, korpi1999supernova}). Instead, if we take the alternate approach of assuming a typical energy of $10^{51}$ erg and estimate the age of a presumed SNR, we get a range of lifetimes within $\sim10-10^3$ years, making it very unlikely for an unreported relatively close and recent supernova to have been responsible for the formation of such a structure.

Furthermore, we note that the lack of any consistent and extended shock tracer emission also makes this scenario very unlikely.

\subsection{Evidence for jet-ISM interactions} \label{sec:jetISM}

We have analyzed the molecular emission in the fields surrounding the BHXBs
MAXI J1348$-$630 and MAXI J1820$+$070, detecting only the $^{13}$CO molecule in the MAXI J1348--630 field. These results are consistent with wide-field CO surveys of the Galaxy \citep{taylor2003canadian}, which trace the distribution of molecular gas over large scales. These surveys show that while MAXI J1348$-$630 lies within a region where CO seems to be quite abundant, MAXI J1820$+$070 appears to lie in a region noticeably devoid of CO emission. Further, the non-detection of any bright molecular emission around MAXI J1820$+$070 is consistent with radio jet motion modeling. In this updated modeling work,  it is shown that a constant density ISM (at values much lower than $n_{\text{crit}}$ typical for the molecular tracers targeted in these observations), rather than an abrupt impact with a cavity wall, can better reproduce the jet dynamics \citep{carotenuto2024constraining, savard2025relativistic}.

For the MAXI J1348--630 field, we interpret the $^{13}$CO emission identified as being physically associated with the BHXB source, based on the following lines of evidence:
\begin{itemize}
\item The morphology of the detected $^{13}$CO emission traces the size and location of the suspected jet blown cavity (Fig.~\ref{fig:main}; \citealt{carotenuto2022modelling}).
\item The brightest detected $^{13}$CO emission coincides with the location of the suspected jet impact sites at the edges of the cavity (Figs.~\ref{fig:spectra} \& \ref{fig:features}).
\item The spectra of the detected $^{13}$CO emission displays characteristics of complex motion (double peaked structure, velocity shifts between regions) driven by collisions in gas clouds (Fig.~\ref{fig:spectra}).
\item The kinematic properties of the detected $^{13}$CO emission show evidence of displaced gas at the suspected cavity walls (Fig.~\ref{fig:PV}).
\item The central velocities of the detected $^{13}$CO emission lines (Fig.~\ref{fig:spectra}) yield kinematic distances consistent with the distances to the central BHXB determined through other independent methods.
\end{itemize}

\subsection{Deriving jet properties}
\label{sec:jetprop}
For MAXI J1348--630 we have identified a structure that is consistent with a jet-driven cavity surrounding the BHXB system. Using a calorimetric approach, we can compute the power that the jet needs to transfer into the local ISM to form and maintain such a structure. Following the formalism introduced by \cite{kaiser1997self} (further developed in Appendix \ref{sec:jetpower}) we can compute the total jet power ($Q_j$; averaged over the jet's lifetime) as a function of ISM impact site characteristics, 
\begin{equation}
    Q_j=\left(\frac{5}{3}\right)^3 \frac{\rho_0}{C_1^5}L_j^2 v^3.
\end{equation}
Here, $L_j$ represents the length of the jet path, $\rho_0$ represents the average density of the impact site, $C_1$ is a factor that depends on the adiabatic indices of the gas in the jet, cavity, and external medium and the opening angle of the jet ($\phi$), and $v$ is the velocity of the shocked gas at the interaction site. 

\begin{figure}[t!]
    \centering
    \includegraphics[width=\linewidth]{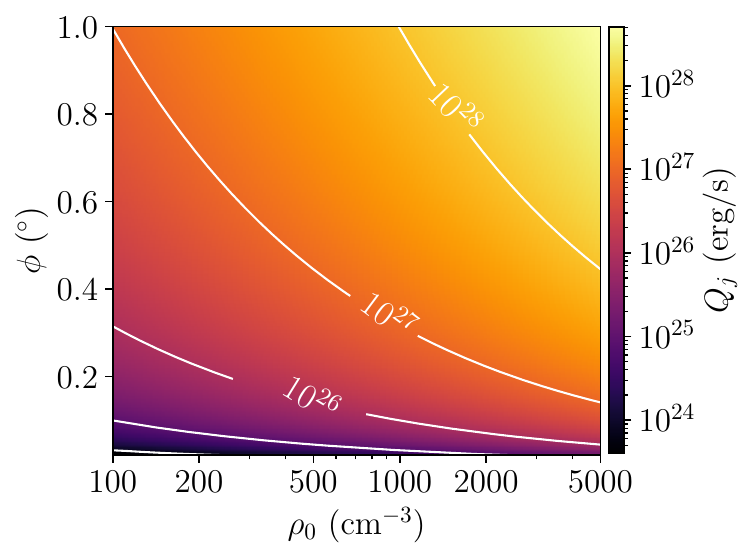}
    \caption{Calorimetric life-time averaged jet power estimates for varying values of impact site ISM density ($\rho_0$) and jet opening angle ($\phi$). We show the density range constrained by the \textsc{radex} modelling and opening angle range below $1^\circ$. This set of results assumes $D=2.2$kpc and $i=50^\circ$. The jet power estimations range from $10^{25}-10^{28}$erg s$^{-1}$.}
    \label{fig:power}
\end{figure}

To estimate $L_j$, we will take the average angular separation between the source and the centroid of the impact sites emission ($\theta\sim[23\pm 1]''$) and explore a range of inclinations with respect to the line of sight ($i\sim[20-80]^\circ$) constrained by previous observations of disk inclination and jet motion \citep{anczarski2020x,carotenuto2021black, cooper2025joint}:
\begin{equation}
    L_j=1.5\times10^{16} (D/\text{kpc})(\theta/\text{rad}) (\sin i)^{-1}  \text{cm,}
\end{equation}
where $D$ is the distance to the source.  Based on the range of our distances and inclination angles, we find that $Q_j$ scaling only leads to factor of 0.3 to 14 changes. Therefore, in Fig. \ref{fig:power} we present results for $D = 2.2$ kpc and $i=50^\circ$, which is sufficient for order of magnitude interpretations.

For the $C_1$ parameter, we take the adiabatic indices $\Gamma_j=\Gamma_x=\Gamma_c=5/3$ and leave the opening angle $\phi$ as a free parameter, that must be lower than 1 degree as in \citealt{carotenuto2022modelling}). 

We estimate the density based on our $^{13}$CO observations together with \textsc{radex} modelling non-local thermal equilibrium (LTE) molecular radiative transfer in an isothermal homogeneous medium \citep{van2007computer}. Since we lack $^{12}$CO observations at the same transition to accurately estimate the optical depth, so we explore a range of situations using constant excitation temperatures for the emission region around 4--6 K (as often found for CO in low density regions; \citealt{mazumdar2021high}) and estimating column densities with the method described in Appendix \ref{sec:columndens}. This provides a range of column densities between $10^{14}-10^{16}$ cm$^{-2}$ for $^{13}$CO. As a consistency check, we assess the average number density of molecular hydrogen in our maps. We transform the column density using the ratio [$^{13}$CO/H$_2$]$=(5.6\pm2.5)\times10^{-5}$ \citep{sofue2020co} and find the resulting values to be in agreement (within one order of magnitude) with those upper limits set by previous X-ray observations of the source ($\sim 10^{21}-10^{22}$ cm$^{-2}$; \citealt{bhowmick2022properties}). With the H$_2$ column densities, we find the observed cloud to have a mass between $(1.6-3.0)(D/2.2$ $\text{kpc})^2M_\odot$ (refer to Appendix \ref{sec:mass} for the specific procedure), reasonable within sub-pc scales. Then, to recover our observed $T_R$ values ($\sim 0.5$ K) through \textsc{radex}, we find that $\rho_0$ can take values between $\rho_0 \sim 100-5000$ cm$^{-3}$. We note that $n_{\text{crit}} \sim 3000$ cm$^{-3}$ for $^{13}$CO, so a significant part of these values will correspond to densities below that threshold. In those cases, LTE cannot be assumed, further justifying our use of \textsc{radex} to account for non-LTE scenarios.
We take $v$ to be the FWHM of spectral features A and B (around $3\,{\rm km\,s}^{-1}$, Fig. \ref{fig:spectra} region 1a,1b,1c).

Fig. \ref{fig:power} displays the resulting lifetime-averaged jet power for different values of distance, $\rho_0$, and $\phi$.
We constrain the energy carried in the jets to be $10^{37}-10^{40}$ erg over a lifetime of a few times $10^4$ years, resulting in a life-time averaged jet power of $Q_j\sim10^{25}-10^{28}$ erg s$^{-1}$.
This value is lower than the typical jet powers computed for the BHXB population ($10^{36}-10^{38}\, {\rm erg\,s}^{-1}$), and lower than the energetics estimates derived from jet motion modeling ($\log_{10}(E_0/\text{erg})\sim 43.7-44.5$; \citealt{zdziarski2023no}) by several orders of magnitude. The disagreement between these values could result from assumptions in the calorimetry calculations (e.g., assuming a homogeneous expansion) or be an indication that the source has spent the majority of its lifetime in a less active state, where powerful jets are not actively produced. Given the transient nature of the source, the latter is likely a significant issue.

To get a different look at the timescales involved in the jet-ISM interaction, we can reconcile the theoretical jet power with the lifetime averaged measurement we get through the calorimetry method. 
Assuming MAXI J1348$-$630 has spin-powered jets, the maximum total jet power available to the source during the launching of the jet ejecta should follow \citep{zdziarski2023no},
\begin{equation}
    P_j \sim 6.3 a_*^2\left(\frac{D}{2.2kpc}\right)^2\left(\frac{\epsilon}{0.05}\right)^{-1}10^{39}\frac{\rm erg}{s}
    \label{eq:maxpower}
\end{equation}
where $a_*$ is the dimensionless spin parameter and $\epsilon$ is the radiative efficiency of the black hole. These parameters are not strictly independent, but their relationship is not well understood. Typically, $a_*$ goes from 0 for a Schwarzschild BH to 1 for a maximally rotating Kerr BH. Within that interval, $\epsilon$ takes values between 0.057 and 0.42. From this last expression we see that physically plausible $P_j$ estimates for a rapidly rotating BH would be  $\sim10^{39}$ erg s$^{-1}$. 

To make an accurate comparison to the life-time averaged jet power calculated through the calorimetry process ($Q_j$) above, we need to take into account the time when the BHXB is in an outburst state vs. in a quiescence state by setting a duty cycle. Assuming that energy is only injected into the ISM during outburst ($E_{\rm TOT}=P_j \cdot t_{\rm ON}$, where time in outburst is denoted by $t_{\rm ON}$), and that the jets remain inactive during quiescence (denoted by $t_{\rm OFF}$), 

\begin{equation}
    Q_j\sim\frac{E_{\rm TOT}}{t_{\rm ON}+t_{\rm OFF}}\sim\eta P_j
\end{equation}
where $\eta=\frac{t_{\rm ON}}{t_{\rm ON}+t_{\rm OFF}}$ is the fraction of the total lifetime where the jet remains active. Comparing our time-averaged calorimetric jet power ($Q_j$) to the theoretical jet power ($P_j$) suggests $\eta\sim10^{-10}-10^{-13}$. This suggests again that the jet has not been actively depositing significant amounts of energy for the majority of its lifetime, but rather likely depositing energy in the form of short-lived jet ejection episodes during infrequent outburst periods. For instance, assuming a total lifetime of the source as an X-ray binary of around 1 Gyr, the jet would have been active between days and years, which is very consistent with the previously observed discrete ejecta emitted over short timescales. 
This would also imply that the transient ejecta feedback into the ISM would be minimal compared to that found for persistent jets ($10^{37}-10^{39}$ erg s$^{-1}$; \citealt{sell2015shell,motta2025meerkat,tetarenko2020jet}).

\subsection{Formation timescales and lifetimes of jet-ISM interaction zones}

Given the structure we observe and taking its physical size ($\sim 0.3-1$pc), if the whole structure had been blown since the last outburst given that the observations were taken $\sim 4$ years post-outburst, the required velocity for the expanding shock front would be around $0.2c-0.7c$, which is completely inconsistent with the dispersion velocities that we observe ($^{13}$CO linewidths $\sim 3$ km s$^{-1}$). This hints towards a previous history of outbursts from the source that might have progressively built the cavity-like structure.
  
We note that the line intensities we find around MAXI J1348--630 are much lower than detected around persistent BHXB sources previously \citep{tetarenko2018mapping,tetarenko2020jet}, which is consistent with the fact that energy could be injected into these structures over short timescales.

However, future monitoring of this and similar jet impact sites post-outburst for transient BHXBs would be needed to constrain how long-lived these molecular features could be.

\section{Summary}
In this study, we investigate the potential interaction between the jets of the BHXBs MAXI J1348--630 and MAXI J1820$+$070 and their surrounding ISM environment. We identify a structure consistent with a jet-driven cavity near MAXI J1348$-$630 and estimate the lifetime averaged jet power required to sustain such a feature using molecular line observations and calorimetry. By applying the formalism of \cite{kaiser1997self} and incorporating observational constraints on jet length, density, and velocity, we compute the  lifetime averaged jet power while exploring the effects of key parameters, such as the jet opening angle and the ISM density. We find that producing physically plausible jet power values consistent with theoretical limits \citep{zdziarski2023no} and previous outburst observations requires extremely rare outbursts where the jets primarily deposit energy into the ISM in the form of short-lived energetic jet ejecta. It therefore also suggests that transient ejecta play a minimal role in ISM feedback when compared to the values found in persistent jets.

We find no evidence for jet-ISM interactions in the case of MAXI J1820$+$070, which is consistent with a jet decelerating through an ISM of constant density much lower than $n_{\text{crit}}$ for the target tracing molecules of this work (as suggested in \citealt{carotenuto2024constraining, savard2025relativistic}).

Our findings reinforce the suitability of the calorimetric approach to estimate jet power, for the first time performing these experiments on transient BHXB sources. 
This method stands as a powerful tool to accurately and efficiently constrain jet powers across the BHXB population using molecular line emission from jet-ISM interaction sites.

\begin{acknowledgments}
The authors thank the anonymous reviewer for their insightful comments, which improved the quality of this manuscript. PB--C and AJT acknowledge that this research was undertaken thanks to funding from the Canada Research Chairs Program and the support of the Natural Sciences and Engineering Research Council of Canada (NSERC; funding reference number RGPIN--2024--04458). ER acknowledges the support of NSERC, funding reference number RGPIN-2022-03499. DMR is supported by Tamkeen under the NYU Abu Dhabi Research Institute grant CASS. GRS acknowledges support from an NSERC Discovery grant (RGPIN-2021-0400). This paper makes use of the following ALMA data: ADS/JAO.ALMA\#2023.1.00983.S. ALMA is a partnership of ESO (representing its member states), NSF (USA) and NINS (Japan), together with NRC (Canada), NSTC and ASIAA (Taiwan), and KASI (Republic of Korea), in cooperation with the Republic of Chile. The Joint ALMA Observatory is operated by ESO, AUI/NRAO and NAOJ. The National Radio Astronomy Observatory is a facility of the National Science Foundation operated under cooperative agreement by Associated Universities, Inc.
\end{acknowledgments}

%
\facilities{ALMA(12m array and ACA)}

\software{\textsc{casa} \citep{bean2022casa}, \textsc{casa} ALMA pipeline \citep{2023PASP..135g4501H}, PHANGS--ALMA pipeline \citep{leroy2021phangs}, \textsc{astropy} \citep{astropy:2013,astropy:2018, astropy:2022}, \textsc{spectralCube} \citep{2019zndo...2573901G}, \textsc{PVextractor} \citep{2016ascl.soft08010G}, \textsc{radex} \citep{van2007computer}. }


\bibliography{sample7}{}
\bibliographystyle{aasjournalv7}

\appendix
\section{Other molecular emission along the line of sight}\label{sec:otheremiss}
\begin{figure*}[t!]
    \centering
    \begin{subfigure}{0.329\textwidth}
        \centering
        \includegraphics[width=\textwidth]{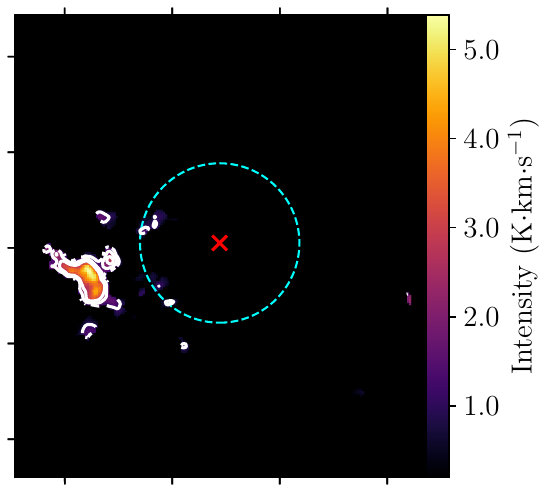} 
    \end{subfigure}
    \hfill
    \begin{subfigure}{0.329\textwidth}
        \centering
        \includegraphics[width=\textwidth]{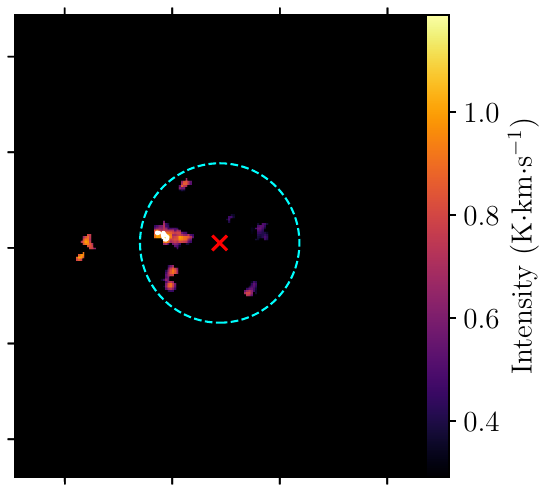} 
    \end{subfigure}
    \hfill
    \begin{subfigure}{0.329\textwidth}
        \centering
        \includegraphics[width=\textwidth]{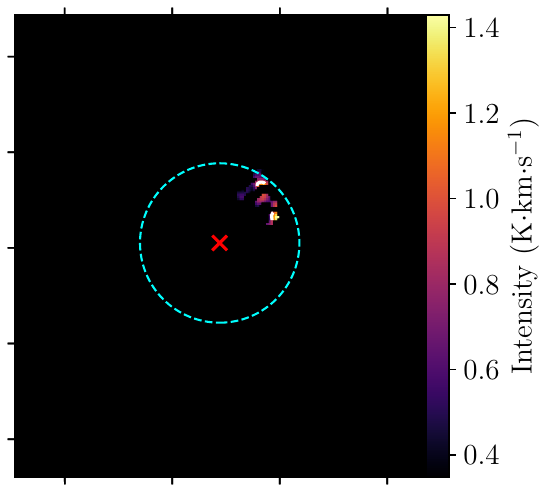} 
    \end{subfigure}

    \caption{Integrated and $4\sigma$ filtered intensity maps of other $^{13}$CO ($J=1-0$) emission features unrelated to MAXI J1348$-$630 or irrelevant to the calorimetric approach along the line of sight. We present the radius at which the ejecta decelerate due to a presumed jet-ISM interaction (dashed cyan lines) and source location (red cross) in the same fashion as previous figures. The {\it left} panel displays the emission between $v=(-63,-58)$ km s$^{-1}$ (Feature E in Figure~\ref{fig:spectra}), the {\it center} panel corresponds to the emission between $v=(-10,-6)$ km s$^{-1}$ (Feature C in Figure~\ref{fig:spectra}), and the {\it right} panel corresponds to the emission between $v=(47,51)$ km s$^{-1}$ (Feature D in Figure~\ref{fig:spectra}). Contours correspond to [1.05, 1.75, 2.45]K km s$^{-1}$. We omit RA and DEC labels for better visualization, but image orientation remains the same as previous figures.}
    \label{fig:array}
\end{figure*}

We have found extended emission at different velocity peaks alongside our line of sight to MAXI J1348$-$630. We have discarded this emission as not being related to the BHXB jet-ISM interaction, but we present it in this section for further assessment (Figure~\ref{fig:array}). It is likely that these molecular line emission regions originate in unrelated molecular clouds, as their central velocities are well within the expected velocity range for molecular emission in the Galactic plane along this line of sight \citep{dame2001milky}. Furthermore, their line profiles are single peaked and reasonably symmetric, which would not be strongly suggestive of a complex kinematics environment consistent with jet impacts.

\section{Integrated intensity maps of all molecular tracers}
\begin{figure*}
\caption{Integrated emission for all tracers in both fields.}
    \centering
    \begin{subfigure}{\linewidth}
        \centering
        \includegraphics[width=\linewidth]{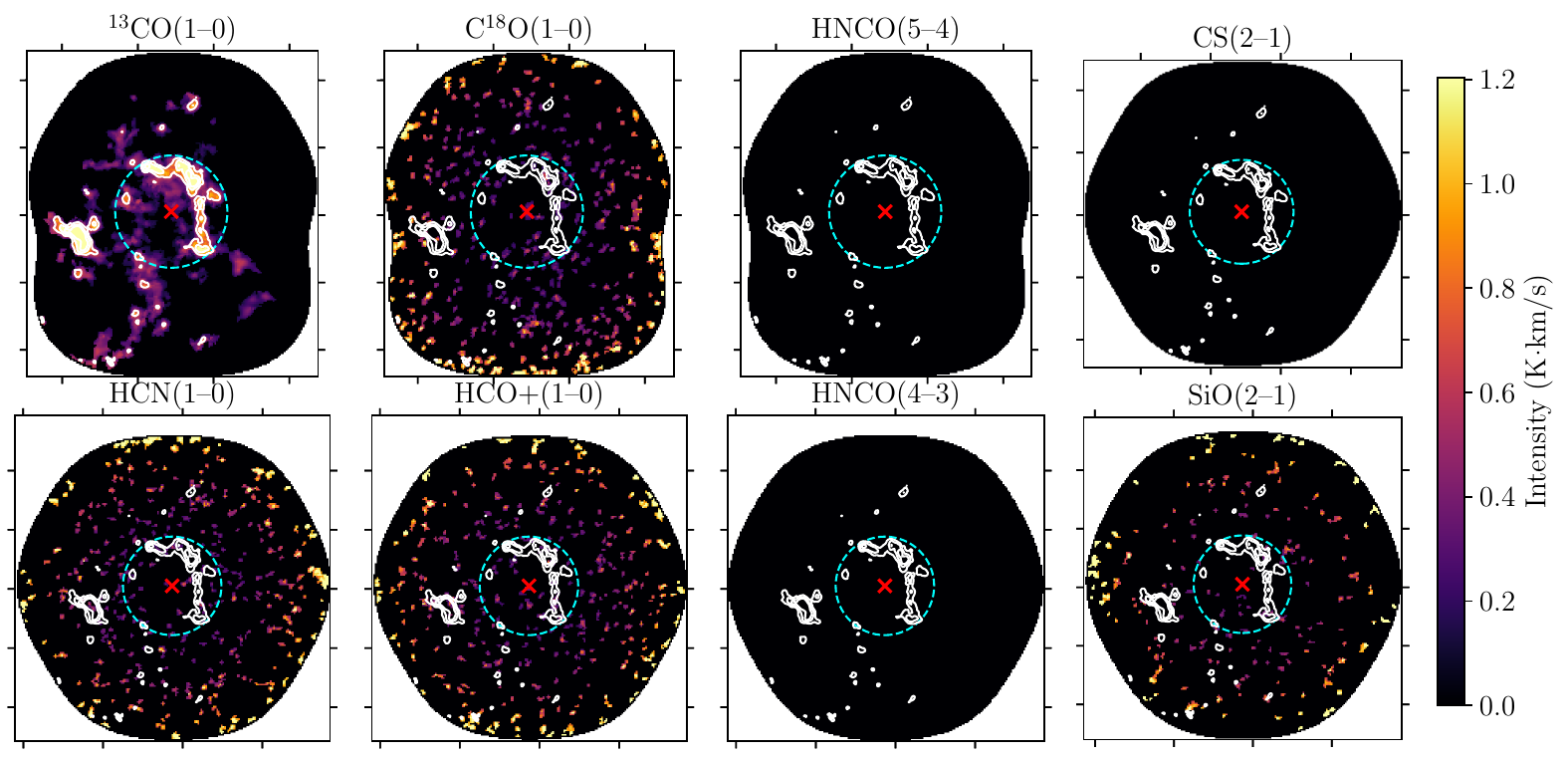}
        \caption{Integrated intensity maps of all the molecular tracers sampled in our ALMA observations for the MAXI J1348$-$630 field. The BHXB source location is marked with a red cross in all panels. The detected $^{13}$CO ($J=1-0$) emission contours (levels: [$1.5,2.5,3.5$] K km s$^{-1}$) are also shown in all panels to contextualize the maps, together with the apparent radius at which the ejecta decelerate due to a presumed jet-ISM interaction (dashed cyan line). We do not find significant or spatially coincident detections of any molecular tracers sampled beyond $^{13}$CO ($J=1-0$) in this field.}
        \label{fig:lines1348}
    \end{subfigure}
    \vspace{1em}  
    \begin{subfigure}{\linewidth}
        \centering
        \includegraphics[width=\linewidth]{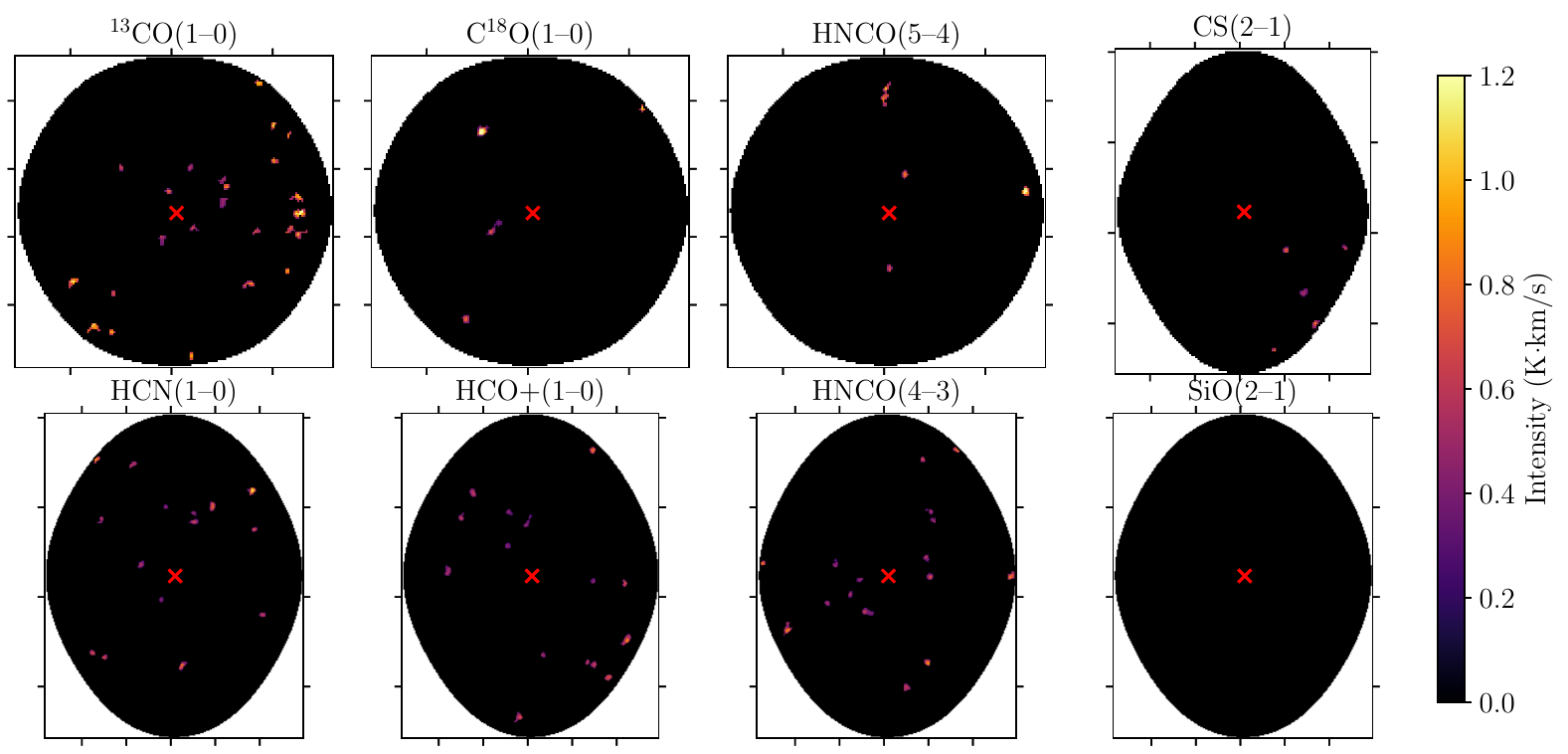}
        \caption{Integrated intensity maps of all the molecular tracers sampled in our ALMA observations for the MAXI J1820$+$070 field. The BHXB source location is marked with a red cross in all panels. We do not find significant detections of any molecular tracers sampled in this field.}
        \label{fig:lines1820}
    \end{subfigure}
\end{figure*}

In this section, we show integrated emission maps for all molecular tracers sampled in our ALMA observations. For MAXI J1348$-$630, we do not detect any significant emission beyond the density tracer $^{13}$CO (Fig.~\ref{fig:lines1348}). For MAXI J1820$+$070, we do not detect significant emission in any of the molecular tracers sampled (Fig.~\ref{fig:lines1820}). For these images we have lowered the detection threshold to $3\sigma$.

\section{ALMA mosaic field and noise map}
\begin{figure*}[h]
    \centering
    \begin{minipage}{0.4\textwidth}
        \centering
        \includegraphics[width=\textwidth]{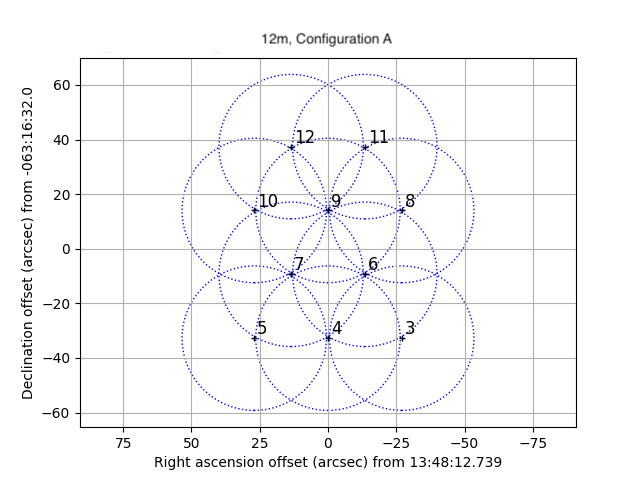}
    \end{minipage}
    \begin{minipage}{0.4\textwidth}
        \centering
        \includegraphics[width=\textwidth]{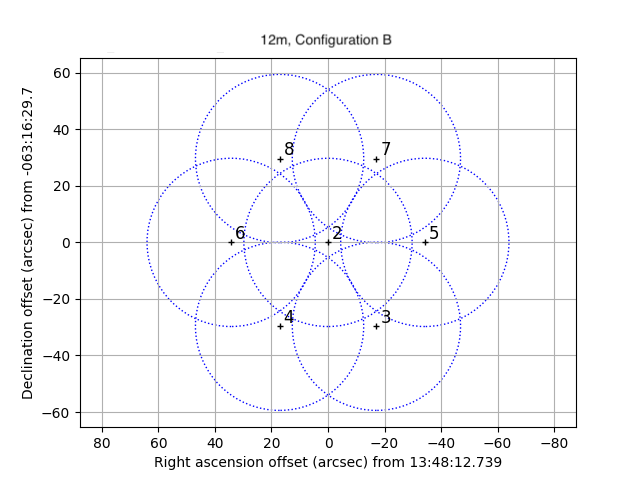}
    \end{minipage}
    \vspace{0.2cm}
    \begin{minipage}{0.4\textwidth}
        \centering
        \includegraphics[width=\textwidth]{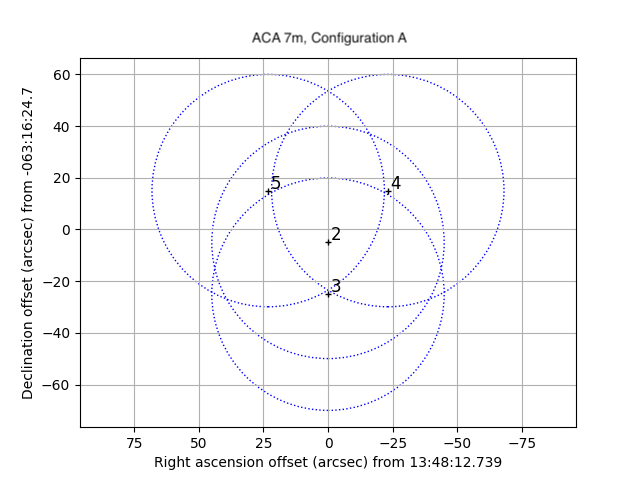}
    \end{minipage}
    \begin{minipage}{0.4\textwidth}
        \centering
        \includegraphics[width=\textwidth]{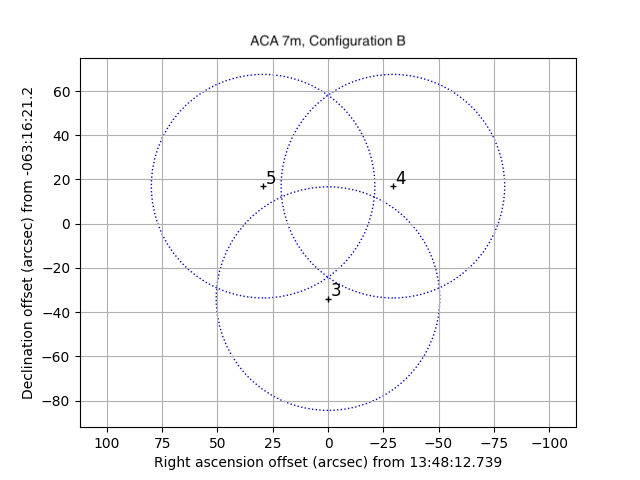}
    \end{minipage}
    \caption{Mosaic maps showing the individual pointings for both the ACA and 12 m arrays in the two spectral setups (labeled as A and B).}
    \label{fig:mosaic}
\end{figure*}

\begin{figure*}[h]
    \centering
    \includegraphics[width=0.5\linewidth]{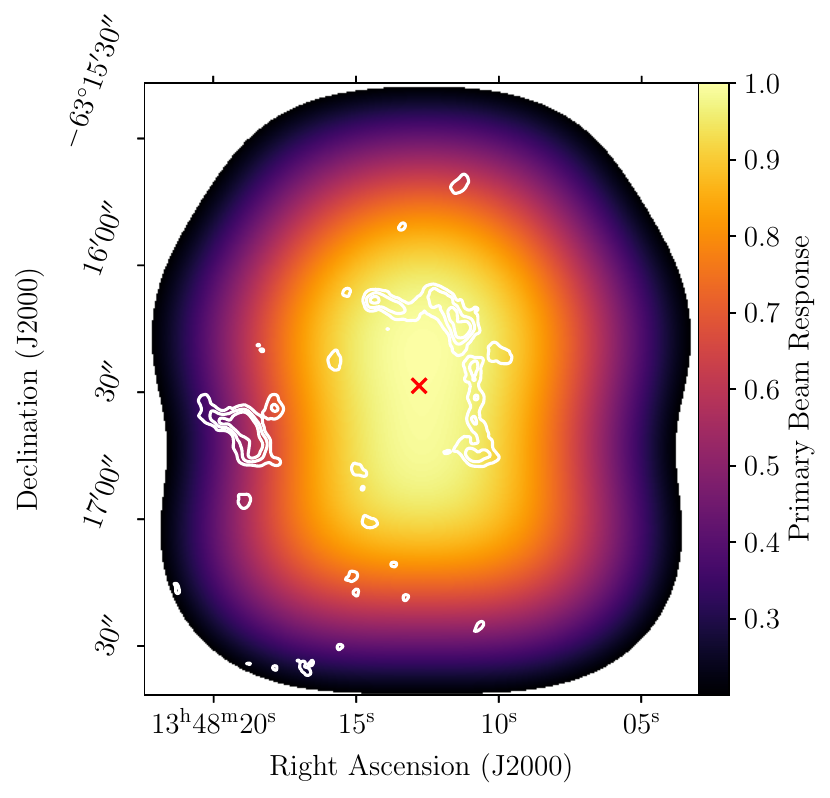}
    \caption{Primary beam response map for the $^{13}$CO($1-0$) observation. The central BHXB position is marked with a red cross. The contours represent the molecular emission from the integrated intensity map presented in Figure~\ref{fig:main} with levels of [$1.5,2.5,3.5$] K km s$^{-1}$. }
    \label{fig:response}
\end{figure*}

For the observations presented in this work (those of MAXI J1348$-$630), we used two separate spectral setups, as noted in Table~\ref{tab:config}. We present the mosaic maps, with each of the pointings for the 12m and ACA arrays in Figure~\ref{fig:mosaic}. 
We also show the primary beam response for the combination of ACA and 12 m data in Figure~\ref{fig:response}.

\section{Column density estimation}\label{sec:columndens}
To derive a column density value for every pixel in our molecular emission map, we follow the procedure of \cite{wilson2009tools}. We assume that $^{13}$CO represents optically thin emission with an excitation temperature ($T_{\rm ex}$) equivalent to that of an optically thick species like $^{12}$CO. The intensity of the line can be described by,
\begin{equation}
    I_{\rm line}=(S-I_0)(1-e^{-\tau})
    \label{eq:C1}
\end{equation}
where $S$ is the source function and $\tau$ is the optical depth. We assume that $S$ and $I_0$ can be described by two black body curves at $T_{\rm ex}$ and $T_{\rm bg}=2.73$K, respectively.

The radiation temperature is represented by,
\begin{equation}
    T_R=I_{\rm line}\frac{c^2}{2\nu^2k}
    \label{eq:Tr}
\end{equation}

Substituting in Equation~\ref{eq:C1},
\begin{eqnarray}
    T_R&=&(S-I_0)(1-e^{-\tau})\frac{c^2}{2\nu^2 k}\\\nonumber
    &=&\frac{h\nu}{k}(1-e^{-\tau})\left(C_{\rm ex}-C_{\rm bg}\right)\\\nonumber
\end{eqnarray}
where,

\begin{equation}
    C_{\rm ex}=\frac{1}{\exp[h\nu/kT_{\rm ex}]-1},
\end{equation}

\begin{equation}
    C_{\rm bg}=\frac{1}{\exp[h\nu/kT_{\rm bg}]-1}.
\end{equation}

We take a range of constant $T_{ex}$ values between $4-6$K given the low $T_R$ values we detect in our spectral cube.The optical depth of the region can be derived from rearranging Equation \ref{eq:Tr},

\begin{equation}
    \tau = -\ln \left[ 1 - \frac{kT_R}{h \nu} \left( C_{\rm ex} - C_{\rm bg}\right)^{-1} \right]
\end{equation}

The column density for the $J$th state with frequency $\nu_0$ is represented as \citep{wilson2009tools},
\begin{equation}
    N_J = \frac{8 \pi \nu_0^3}{c^3} \frac{g_l}{g_u} \frac{1}{A_{ul}}   \left(1 - e^{-\frac{h \nu_0}{k T_{\text{\rm ex}}}}\right)^{-1} \int_{v_{\rm min}}^{v_{\rm max}} \tau dv
\end{equation}

where $A_{ul}=1.165\times10^{11}\mu^2\nu_0^3\frac{J+1}{2J+3}$ for the $J+1\rightarrow J$ transition, and $\mu^2=0.122D$ for CO.

Finally, for the total density, we assume local thermal equilibrium (LTE) for a CO molecule ($J+1\rightarrow J$ rotational transition), with a population characterized by $T_{\rm ex}$ yielding,
\begin{equation}
    N_{\rm tot} = N_J \times\frac{Z(T_{\rm ex})}{2J+1}\times e^{\frac{2.65 J(J+1)}{T_{\rm ex}}}.
\end{equation}

Here the 2.65 value is the $hB_e/k = 2.65 K$ ratio specific to $CO$, and the partition function can be expressed as,

\begin{equation}
   Z (T_{\rm ex}) = \sum_{J=0}^{\infty}(2J+1)\times e^{-\frac{2.65 J(J+1)}{T_{\rm ex}}} 
\end{equation}

\section{Computing the mass of the cloud from the column density map}\label{sec:mass}

We compute the mass of the cloud for both the existing distance estimates \citep{chauhan2021measuring,lamer2021giant}) and a [$^{13}$CO/H$_2$] ratio \citep{sofue2020co} using the following approach,
\begin{equation}
    M=\sum_{\rm im} N_{\rm pix}  \theta_{\rm pix}^2 D^2  m_H  \mu,
\end{equation}
where $N_{\rm pix}$ is the column density value of a given pixel, $\theta_{\rm pix}$ is the resolution of the pixel in radians, $D$ is the distance to the source in cm, $m_H$ is the atomic hydrogen mass and $\mu$ is a factor that accounts for the fact that there is 2 hydrogen atoms in every molecule and also extra mass related to helium and heavier elements (typically $2\times1.36$ for molecular clouds \citealt{keilmann2024molecular}). 

\section{Calorimetry method} \label{sec:jetpower}
Following \cite{kaiser1997self}, we can derive the total power of the jet by relating it to a jet-ISM interaction site and its properties. We assume the jets remain pointed in a constant direction (i.e., not precessing) and the impact to happen at a site with a constant $\rho_0$, with a constant jet power averaged over jet lifetime $Q_j$. The length of the jet as a function of time can be expressed as,
\begin{equation}
    L_j=C_1\left(\frac{t}{\tau}\right)^{\frac{3}{5-\beta}},
    \label{eq:Lj}
\end{equation}
where the characteristic timescale is $\tau=(\rho_0/Q_{j})^{1/3}$ and constant terms are represented as,
\begin{equation}
\small
    C_1 = \left[ \frac{C_2}{C_3 \phi^2} \frac{(\Gamma_x + 1)(\Gamma_c - 1)(5 - \beta)^3}{18  \{9 [\Gamma_c + (\Gamma_c - 1) \frac{C_2}{4 \phi^2} ] - 4 - \beta \}} \right]^{\frac{1}{5 - \beta}}
\end{equation}

\begin{equation}
    C_2 = \left[ \frac{(\Gamma_c - 1)(\Gamma_j - 1)}{4 \Gamma_c} + 1 \right]^{\frac{\Gamma_c}{(\Gamma_c - 1)}} \frac{(\Gamma_j + 1)}{(\Gamma_j - 1)}
\end{equation}

\begin{equation}
    C_3 = \frac{\pi}{4 R_{\text{ax}}^2}.
\end{equation}
Here the axial ratio of the jet blown cavity $R_{\rm ax}=\sqrt{\frac{1}{4}\frac{C_2}{\phi^2}}$, $\phi$ is the opening angle of the jet (in radians), and $\Gamma_j$, $\Gamma_x$, and $\Gamma_c$ represent the adiabatic indices of the jet, the jet-blown cavity and the external medium with which the jet is colliding, respectively.

To create and maintain the jet-blown cavity structure in the ISM, the jet would have to carry over its lifetime ($t$) a power that can be expressed by rearranging Equation~\ref{eq:Lj},
\begin{equation}
    Q_j=\rho_0 \left(\frac{L_j}{C_1}\right)^5t^{-3}
    \label{eq:pow1}
\end{equation}
Here the assumption of a constant density medium lets us work with $\beta=0$.

To estimate the jet lifetime, we take the time derivative of Equation~\ref{eq:Lj}, representing the velocity of the shocked gas at the interaction site (again assuming constant density with $\beta=0$):
\begin{equation}
    \dot{L}_j=\frac{3}{5}C_1\left(\frac{Q_{j}}{\rho_0}\right)^{\frac{1}{5}}t^{-\frac{2}{5}}=v.
    \label{eq:ldot}
\end{equation}
\newline
Combining Equations~\ref{eq:Lj} \& \ref{eq:ldot},
\begin{equation}
    t=\frac{3}{5}\left(\frac{L_j}{v}\right),
\end{equation}

Substituting this jet lifetime expression into Equation~\ref{eq:pow1} yields,
\begin{equation}
    Q_j=\left(\frac{5}{3}\right)^3 \frac{\rho_0}{C_1^5}L_j^2 v^3
    \label{eq:time}
\end{equation}



\end{document}